\documentclass[twocolumn]{aastex631}

\usepackage{graphicx}
\usepackage{mathtools}
\usepackage[mathlines]{lineno} 
\usepackage{amsmath,amsthm,amssymb,bbm}
\usepackage{soul}
\usepackage{comment}
\usepackage{cleveref}
\usepackage{CJK}
\usepackage{appendix}
\usepackage{makecell}
\usepackage{multirow}
\usepackage{subfigure}
\def\[#1\]{\begin{equation}\begin{aligned}#1\end{aligned}\end{equation}}
\def\*[#1\]{\begin{align*}#1\end{align*}}

\begin{document}
\begin{CJK}{UTF8}{gbsn}

\title{Candidate Dark Galaxy-2: Validation and Analysis of an Almost Dark Galaxy in the Perseus Cluster}

\correspondingauthor{Dayi (David) Li}
\email{dayi.li@mail.utoronto.ca}

\shorttitle{Validation of an Almost Dark Galaxy}
\shortauthors{Li et al.}

\author[0000-0002-5478-3966]{Dayi (David) Li (李大一)}
\altaffiliation{Data Sciences Institute Doctoral Fellow,}
\altaffiliation{CANSSI Ontario Multi-disciplinary Doctoral Trainee}
\affiliation{Department of Statistical Sciences, University of Toronto, 700 University Avenue, Toronto, ON M5G 1Z5, Canada}
\affiliation{Data Sciences Institute, University of Toronto, 700 University Avenue, Toronto, ON M5G 1Z5, Canada}

\author[0000-0002-7490-5991]{Qing Liu (刘青)}
\affiliation{Leiden Observatory, Leiden University, P.O. Box 9513, 2300 RA Leiden, The Netherlands}

\author[0000-0003-3734-8177]{Gwendolyn M. Eadie}
\affiliation{Department of Statistical Sciences, University of Toronto, 700 University Avenue, Toronto, ON M5G 1Z5, Canada}
\affiliation{Data Sciences Institute, University of Toronto, 700 University Avenue, Toronto, ON M5G 1Z5, Canada}
\affiliation{David A. Dunlap Department of Astronomy \& Astrophysics, University of Toronto, 50 St George St, Toronto, ON M5S 3H4, Canada}

\author[0000-0002-4542-921X]{Roberto G. Abraham}
\affiliation{David A. Dunlap Department of Astronomy \& Astrophysics, University of Toronto, 50 St George St, Toronto, ON M5S 3H4, Canada}
\affiliation{Dunlap Institute for Astronomy \& Astrophysics, University of Toronto, 50 St George St, Toronto, ON M5S 3H4, Canada}

\author[0000-0002-1442-2947]{Francine R. Marleau}
\affiliation{Universit\"at Innsbruck, Institut f\"ur Astro- und Teilchenphysik, Technikerstrasse 25/8, 6020 Innsbruck, Austria}

\author[0000-0001-8762-5772]{William E. Harris}
\affiliation{Department of Physics and Astronomy, McMaster University, Hamilton, ON L8S 4M1, Canada}

\author[0000-0002-8282-9888]{Pieter van Dokkum}
\affiliation{Department of Astronomy, Yale University, New Haven, CT 06511, USA}

\author[0000-0003-2473-0369]{Aaron J. Romanowsky}
\affiliation{Department of Physics \& Astronomy, San Jos\'{e} State University, One Washington Square,
San Jose CA 95192,
USA}
\affiliation{Department of Astronomy \& Astrophysics, University of California Santa Cruz, 1156 High Street, Santa Cruz, CA 95064, USA}

\author[0000-0002-1841-2252]{Shany Danieli}
\affiliation{School of Physics and Astronomy, Tel Aviv University, Tel Aviv 69978, Israel}
\affiliation{Department of Astrophysical Sciences, 4 Ivy Lane, Princeton University, Princeton, NJ 08544, USA}

\author[0000-0003-2541-3744]{Patrick E. Brown}
\affiliation{Department of Statistical Sciences, University of Toronto, 700 University Avenue, Toronto, ON M5G 1Z5, Canada}
\affiliation{Center for Global Health Research, St. Michael's Hospital, 30 Bond Street, Toronto, ON M5B 1W8, Canada}

\author[0000-0002-4133-6884]{Alex Stringer}
\affiliation{Department of Statistics and Actuarial Science, University of Waterloo, 200 University Avenue West, Waterloo, ON N2L 3G1, Canada}



\begin{abstract}
Candidate Dark Galaxy-2 (CDG-2) is a potential dark galaxy consisting of four globular clusters (GCs) in the Perseus cluster, first identified in \cite{li2025poisson} through a sophisticated statistical method. The method searched for over-densities of GCs from a \textit{Hubble Space Telescope} (\textit{HST}) survey targeting Perseus. Using the same \textit{HST} images and the new imaging data from the \textit{Euclid} survey, we report the detection of extremely faint but significant diffuse emission around the four GCs of CDG-2. We thus have exceptionally strong evidence that CDG-2 is a galaxy. This is the first galaxy detected purely through its GC population. Under the conservative assumption that the four GCs make up the entire GC population, preliminary analysis shows that CDG-2 has a total luminosity of $L_{V, \mathrm{gal}}= 6.2\pm{3.0} \times 10^6 L_{\odot}$ and a minimum GC luminosity of $L_{V, \mathrm{GC}}= 1.03\pm{0.2}\times 10^6 L_{\odot}$. Our results indicate that CDG-2 is one of the faintest galaxies having associated GCs, while at least $\sim 16.6\%$ of its light is contained in its GC population. This ratio is likely to be much higher ($\sim 33\%$) if CDG-2 has a canonical GC luminosity function (GCLF). In addition, if the previously observed GC-to-halo mass relations apply to CDG-2, it would have a minimum dark matter halo mass fraction of $99.94\%$ to $99.98\%$. If it has a canonical GCLF, then the dark matter halo mass fraction is  $\gtrsim 99.99\%$. Therefore, CDG-2 may be the most GC dominated galaxy and potentially one of the most dark matter dominated galaxies ever discovered.    
\end{abstract}



\section{Introduction}\label{sec:intro}

Ultra-diffuse galaxies (UDGs) are a class of galaxies that have very low-surface brightness ($g$-band central surface brightness of $\mu_{0,g} > 24.0$ mag arcsec$^{-2}$) but large physical size (effective radius of $R_e > 1.5$~kpc). Significant attention has been focused on UDGs since their initial identification en masse by \cite{vanDokkum2015} using the Dragonfly Telephoto Array \citep{Abraham2014}. \cite{vanDokkum2015} first identified a large number of UDGs in the Coma cluster by their distinctive smooth, diffuse emission. Subsequently, thousands of UDGs were found in various rich galaxy clusters as well as low density field environments \citep[e.g.,][]{Martinez-Delgado2016, Yagi2016,Wittmann2017,Janssens2019,Roman+2019, Danieli_2019, Forbes+2019, Lim2020,Forbes2020,Danieli+2020, marleau_2021, marleau2024}.

Many UDGs identified to date host exceptionally large globular cluster (GC) populations: 5--7 times more GCs on average than typical galaxies with the same luminosity \citep{Peng2016, van_Dokkum_2017, Amorisco2018, Lim2018, Forbes+2020, Danieli2022}. Some of the most extreme GC-to-stellar mass ratios to date were found in NGC~5846-UDG1 \citep{forbes2019ultra, mueller2021dwarf} with $\sim54$ GCs that make up $\sim9.8\%$ of the total stellar mass \citep{Danieli2022}, UGC~9050-Dw1, with $\sim52$ GCs making up $\sim 16\%$ \footnote{This estimate is obtained by assuming $M_{\rm GC}/M_{\odot} = 2\times 10^5N_{\rm GC}$  \citep{Gannon2024UDG}.} of the stellar mass \citep{fielder2023disturbed}, and VLSB-B with $\sim 26$ GCs taking up $\sim 23.7\%$ of the total stellar mass \citep{Lim2020, Toloba2023NGVS, Gannon2024UDG}.

The discovery of numerous UDGs with large GC populations pushes the boundaries of galaxy formation theories, and begs the question: Are UDGs at the bright end of an even more diffuse class of galaxies? These galaxies could be so diffuse that they are almost entirely dominated by dark matter, with the majority of their stellar populations contained in GCs \citep{Li2022}. In other words, could almost dark galaxies exist?

Under the star formation scenario proposed by \cite{Danieli2022} where nearly all stellar populations may have originated from GCs for NGC~5846-UDG1 and UGC~9050-Dw1, it should not be a wild speculation that almost dark galaxies can exist. The discovery of these galaxies would have profound implications, as they would provide evidence for extreme star formation scenarios as opposed to the typical one where stars are born in loose agglomerations and then slowly disperse such as in the Milky Way \citep{Kennicutt_2012}. In addition, further constraints on GC mass loss may be established by these dark galaxies. Moreover, they could be ideal objects to test dark matter models such as the ultra-light axionic dark matter model \citep{Hu2000, Walker2011, Hui2017}. For example, the famous dark matter dominated UDG Dragonfly~44 \citep{vanDokkum2016, Wasserman2019} has been utilized to constrain the ultra-light scalar field mass within the fuzzy dark matter model framework. Additional massive dark galaxies can provide further constraints on these dark matter models \citep{Wasserman2019, Burkert2020}.

Conventional methods of searching for diffuse stellar light to find UDGs \citep[e.g.,][]{vanDokkum2015} will fail at identifying dark galaxies, even if they exist. To this end, \cite{Li2022} proposed a statistical approach that bypasses the search for diffuse stellar light and instead looks for overdensities of GCs that seemingly do not belong to any bright galaxy. GCs would not clump together spatially without sufficient mass (such as dark matter) binding them gravitationally. Thus, the identification of GC spatial overdensities could imply the existence of UDGs/dark galaxies. 

\cite{Li2022} applied their method to GC data obtained from the PIPER survey \citep{Harris2020} --- a \textit{Hubble Space Telescope} (\textit{HST}) imaging program targeting the Perseus cluster. \cite{Li2022} successfully detected $11$ over-densities of GCs, $10$ of which correspond to previously confirmed UDGs with GC number $N_{\text{GC}} \geq 3$. However, a clump of four tightly-grouped GCs with no apparent diffuse emission was also found. This spatial clump of GCs does not belong to any previously known galaxy, and was therefore labeled Candidate Dark Galaxy-1 \citep[CDG-1;][]{Li2022}. A follow-up study based on a deep \textit{HST}/UVIS imaging program \citep{vanDokkum_2024} still did not reveal any significant diffuse component in CDG-1. In addition, \cite{marleau2024} did not find an optical counterpart for a diffuse emission in the \textit{Euclid} Early Release Observations (ERO) either. Thus, the nature of CDG-1 remains uncertain.

Later, \cite{li2025poisson} addressed several shortcomings of the detection method in \cite{Li2022} and proposed an improved statistical model to detect over-densities of GCs that may be associated with UDGs/dark galaxies. The new method includes additional information such as GC color to boost the clustering signals of GCs within UDGs/dark galaxies. Using the new method and the same GC data from the PIPER survey, \cite{li2025poisson} found another tight grouping of three GCs that was missed in \cite{Li2022}. This spatial clump of GCs was labeled Candidate Dark Galaxy-2 (CDG-2; at coordinates $\alpha = 3^{\mathrm{h}}17^{\mathrm{m}}12^{\mathrm{s}}.61, \delta = 41^\circ20'51''.5$).

In this study, we use the same PIPER imaging material and apply the method in \cite{li2025poisson} to a new GC catalog with photometry more sensitive to the faint end completed in \cite{li2024mathpop}. In this updated analysis, a new GC candidate was detected in CDG-2 and the detection signal for CDG-2 increases by almost $10$-fold compared to that in \cite{li2025poisson}. This puts CDG-2 in the same category as CDG-1 in terms of the clustering strength of its constituent GCs. In other words,  the probability that CDG-2 is a clump of GCs arising by chance from the GC population of the inter-galactic medium is extremely low.

We subsequently noted that CDG-2 was observed in two different imaging visits of the PIPER survey. In this work, we stack these two images and find that there is extremely faint but significant diffuse emission around CDG-2. In addition, we utilize the newly released imaging data from the \textit{Euclid} ERO targeting the Perseus cluster \citep{marleau2024}. The \textit{Euclid} data also reveal extremely faint diffuse emission with the same morphology as that from the \textit{HST} data. The existence of diffuse emission from both \textit{HST} and \textit{Euclid} data provides almost definitive evidence that CDG-2 is a galaxy and it is the first one discovered through its GC population. Using the \textit{Euclid} imaging, which is optimized for detecting diffuse structure, we conduct a simple analysis which suggests that at least $16.6\%$ of light in CDG-2 comes from its GC population, while a much higher ratio of $33\%$ is possible if additional but unobserved GCs are present. Thus, CDG-2 may be the galaxy with the most extreme GC stellar light and mass ratios ever discovered.
\section{Data}\label{sec:data}

\begin{figure*}[t]
    \centering
    \includegraphics[width = \textwidth]{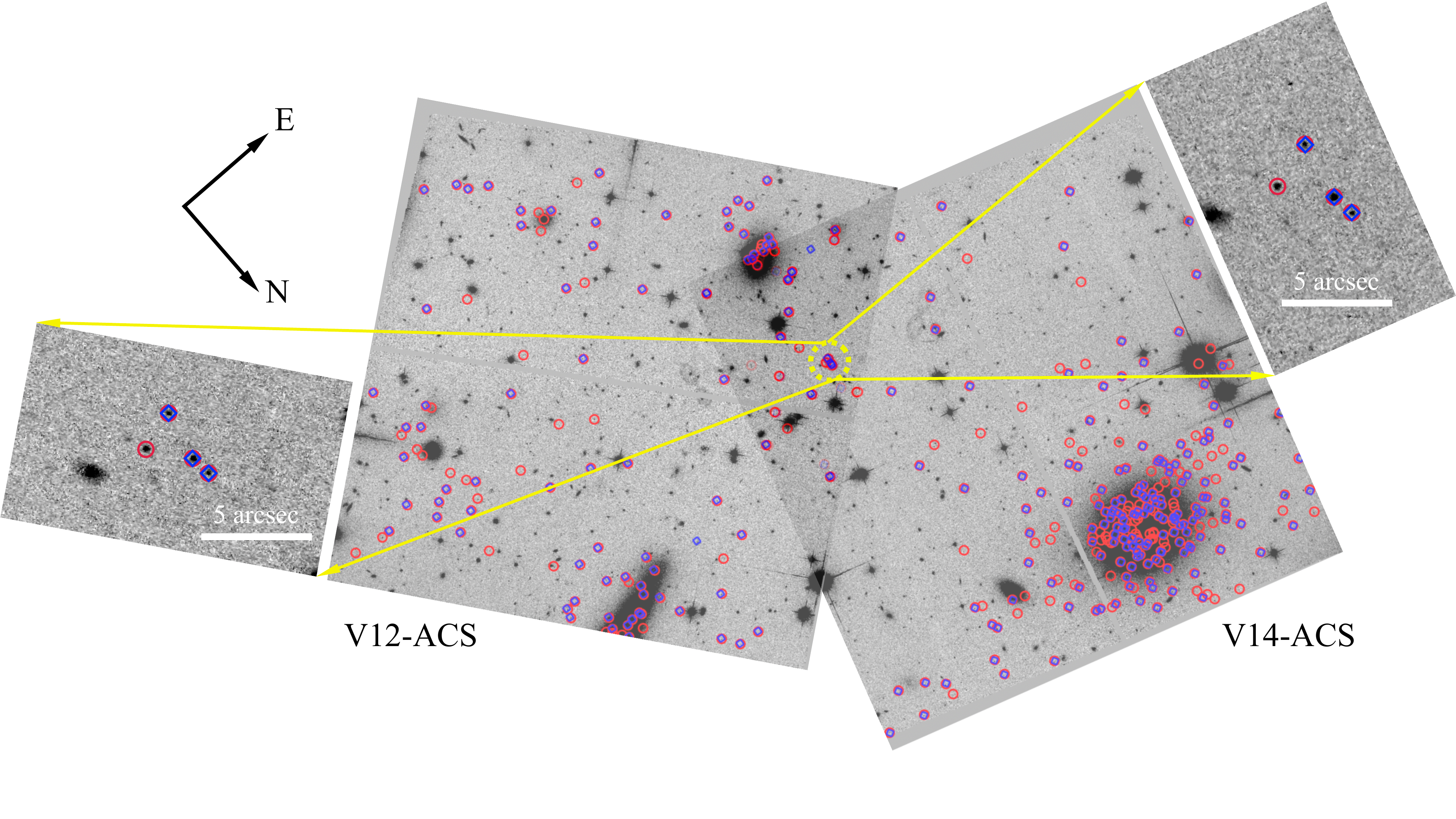}
    \caption{Spatial distributions of GC candidates in the F814W images V12-ACS (left) and V14-ACS (right). Red circles are GC candidates from DOLPHOT while blue diamonds are from DAOPHOT. The GC candidates that constitute CDG-2 from the two images are enlarged and annotated on the side.}
    \label{fig:CDG2_diagram}
\end{figure*}

The GC data were constructed from the Program for Imaging of the PERseus cluster (PIPER; \citealt{Harris2020}) survey. The survey targeted the Perseus galaxy cluster at a distance of $75$ Mpc \citep{Harris2020} and was conducted by the \textit{HST} with its on-board Advanced Camera for Surveys (ACS) and Wide Field Camera 3 (WFC3).

Ten imaging visits of the PIPER survey targeted the outer regions of Perseus, where each visit consists of an imaging pair -- one captured by ACS and the other by WFC3. The ACS images were taken using the F475W and F814W filters, while the parallel WFC3 images used the F475X and F814W filters. Each image is assigned an ID denoting the visit number and the camera, e.g., V6-ACS is the image captured by ACS during the 6th visit.

The majority of GCs appear unresolved (pointlike or near-pointlike) at the distance of Perseus. Therefore, photometric software such as DAOPHOT and DOLPHOT designed for this purpose can be used to construct GC catalogs. In the initial PIPER survey paper \citep{Harris2020}, a GC catalog was constructed via the point source list obtained from DAOPHOT \citep{stetson1987}. Later, in \cite{li2024mathpop}, DOLPHOT \citep{Dolphin_2000, dolphin2016} was used to construct a new GC catalog 
to conduct a GC population study for UDGs in Perseus. The details for GC catalogs construction are provided in \cite{Harris2020, Harris_2023, li2024mathpop}. Briefly, for both catalogs, an initial list was constructed from all objects detected in F814W, and magnitudes were measured through PSF (point spread function) fitting. Objects successfully measured in both filters were retained, and clearly non-stellar objects were removed with 
the \texttt{sharp} and \texttt{chi} parameters generated by DOLPHOT \citep[see][for a more extended description]{Harris2020,Harris_2023}. Although the the final measurements from both DOLPHOT and DAOPHOT were in close agreement, it was felt that an independent check on the previous data would be of some value \citep[for a much more exhaustive comparison of the codes in practice, see][]{monelli+2010}.

A careful manual selection of the DOLPHOT point source list was also performed to finalize the GC catalog. The final GC catalogs were determined through the (extinction corrected) color--magnitude diagram (CMD): in \cite{Harris2020}, sources with magnitudes $22.0 \leq \mathrm{F814W} \leq 25.5$~mag and colors $1.0 \leq \mathrm{F475W} - \mathrm{F814W} \leq 2.4$~mag were deemed to be likely GCs; in \cite{li2024mathpop}, a different method was used to obtain a probabilistic GC catalog. In this study, for simplicity, we consider GC selection criteria of $22.0 \leq \mathrm{F814W} \leq 25.75$~mag and colors $1.0 \leq \mathrm{F475W} - \mathrm{F814W} \leq 2.4$~mag based on the point source list from \cite{li2024mathpop}. The faint limit of $25.75$~mag corresponds to the $50\%$ completeness fraction obtained from an artificial star test for DOLPHOT \citep[see][for more details]{li2024mathpop}, while the canonical GC luminosity function (GCLF) turn-over point is $M_{\rm TO} \sim 26.3$~mag at the distance of Perseus \citep{Harris2020, janssens_2024}.

With the recent public release of the \textit{Euclid} Early Release Observations (ERO) targeting the Perseus cluster \citep{marleau2024}, we also utilize the \textit{Euclid} ERO data in this work. While the PIPER survey from \textit{HST} is optimized to detect point sources such as unresolved GCs in Perseus, \textit{Euclid} is better suited for detecting and analyzing extremely diffuse emission that would validate the nature of CDG-2 \citep{cuillandre2024}.

CDG-2 was detected in the images V12-ACS and V14-ACS. The spatial distributions of GC candidates in DOLPHOT and DAOPHOT catalogs from both images are shown in Figure \ref{fig:CDG2_diagram}. The GC candidates that constitute CDG-2 from both images are highlighted in the zoomed-in images on the side of Figure \ref{fig:CDG2_diagram}. Since a portion of large GCs at the distance of the Perseus cluster ($75$~Mpc) are partially resolved and not exactly starlike, the point source selection criteria under DOLPHOT took this into account, as it did with the previous study of \citet{Harris2020}. Additionally, the faint limit of the DOLPHOT GC catalog was deliberately set to be slightly deeper (a lower SNR threshold) with a less stringent point source selection criteria (based on \texttt{sharp} and \texttt{chi} parameters) than those used in the previous study, and thus more GC candidates are present in the present DOLPHOT catalog.

Although the less stringent GC selection criteria means there are more contaminant objects, it also means that fainter or marginally resolved GC candidates previously excluded in the DAOPHOT catalog are now included in the present GC catalog. A direct implication here is that our DOLPHOT catalog contains an additional GC candidate (see Figure \ref{fig:CDG2_diagram}) in close vicinity of the original three GC candidates from \citet{Harris2020} in CDG-2. Table \ref{tab:GC_candidates} shows the (averaged) properties of the four GC candidates from both V12-ACS and V14-ACS measured in the present study. The half-light radii $r_h$ were measured with the ISHAPE profile-fitting code \citep{larsen1999} as described in \citet{Harris2020} and \citet{Li2022}.
The additional GC candidate from the DOLPHOT catalog is CDG-2-GCC1. This GC candidate was missed in the previous catalog most likely because it is marginally resolved (see Figure \ref{fig:CDG2_diagram}) and did not pass the point source selection thresholds that were adopted previously. Close visual inspection of CDG-2-GCC1 suggests that it has the morphology expected from a GC at the distance of Perseus, and it also has the magnitude and color of a typical GC. Additionally, we compared the magnitude measurements from DOLPHOT to that from DAOPHOT for the three GC candidates that are present in both catalogs, and found the measurements are consistent within the measurement uncertainty (a few hundredths of magnitude difference). Based on the results in Table \ref{tab:GC_candidates}, all four of the GC candidates have luminosities, intrinsic colors, and half-light radii that are consistent with identification as GCs.

\begin{deluxetable*}{lcccccc}[ht]

\tablecaption{Data for the globular cluster candidates (GCC) in \mbox{CDG-2} and \mbox{CDG-1} \citep[cf. Table 3 in][]{Li2022}. First column contains the ID of the four GC candidates. R.A. and Dec. provide the coordinates of the GCs. F$814$W$_0$ gives the extinction-corrected magnitude in the F$814$W for the GCs obtained by DOLPHOT. The fifth column shows the color in F$475$W$-$F$814$W. $r_h$ is the half-light radius of each GC while the last column gives the measured luminosity of each GC in $I$-band. Note that the photometry for CDG-2 GCCs are taken as the average of the measurements from V12-ACS and V14-ACS.}\label{tab:GC_candidates}
\tablenum{1}

\tablehead{\colhead{ID} & \colhead{R.A.} & \colhead{Dec.} & \colhead{F814W$_0$} & \colhead{(F475W$-$F814W)$_0$} & \colhead{$r_h$} & \colhead{$\log(L_I/L_{\odot})$}  \\
\colhead{} & \colhead{(J2000)} & \colhead{(J2000)} & \colhead{(Vega-mag)} & \colhead{(Vega-mag)} &  \colhead{(pc)} & \colhead{}  }
\startdata
CDG-2-GCC1 & $3^{\mathrm{h}} 17^{\mathrm{m}} 12^{\mathrm{s}}.50$ & $+41^\circ 20' 51''.00$ & $24.582 \pm 0.0490$ & $1.296 \pm 0.103$ & 2.9 & 5.62  \\ 	
CDG-2-GCC2 & $3^{\mathrm{h}} 17^{\mathrm{m}} 12^{\mathrm{s}}.66$ & $+41^\circ 20' 50''.58$ & $25.193 \pm 0.0725$  & $1.415 \pm 0.159$  & 5.6 & 5.35  \\ 	
CDG-2-GCC3 & $3^{\mathrm{h}} 17^{\mathrm{m}} 12^{\mathrm{s}}.62$ & $+41^\circ 20' 52''.79$ & $24.025 \pm 0.0341$  & $1.423 \pm 0.077$ & $\lesssim 2$ & 5.80  \\ 	
CDG-2-GCC4 & $3^{\mathrm{h}} 17^{\mathrm{m}} 12^{\mathrm{s}}.63$ & $+41^\circ 20' 53''.68$ & $25.183 \pm 0.0720$  & $1.335 \pm 0.153$ & $\lesssim 2$ & 5.35  \\
\tableline
CDG-1-GCC1 & $3^{\mathrm{h}}  18^{\mathrm{m}}  12^{\mathrm{s}}.23$ & $+41^\circ 45' 58''.03$ &	24.709 $\pm$ 0.035 & 1.445 $\pm$ 0.088 & 4.3 & 5.49 \\ 	
CDG-1-GCC2 & $3^{\mathrm{h}}  18^{\mathrm{m}}  12^{\mathrm{s}}.43$ & $+41^\circ 45' 59''.55$ & 	23.535 $\pm$ 0.046 & 1.606 $\pm$ 0.066 & 6.7 & 5.97  \\ 	
CDG-1-GCC3 & $3^{\mathrm{h}}  18^{\mathrm{m}}  12^{\mathrm{s}}.06$ & $+41^\circ 45' 57''.60$ & 	24.926 $\pm$ 0.050 & 1.446 $\pm$ 0.098 & 6.7 & 5.41  \\ 	
CDG-1-GCC4 & $3^{\mathrm{h}}  18^{\mathrm{m}}  12^{\mathrm{s}}.29$ & $+41^\circ 45' 57''.23$ &	24.773 $\pm$ 0.039 & 1.684 $\pm$ 0.076 & 6.5 & 5.48 \\
\enddata

\end{deluxetable*}

\begin{figure*}[t]
    \centering
\includegraphics[width = 0.86\textwidth]{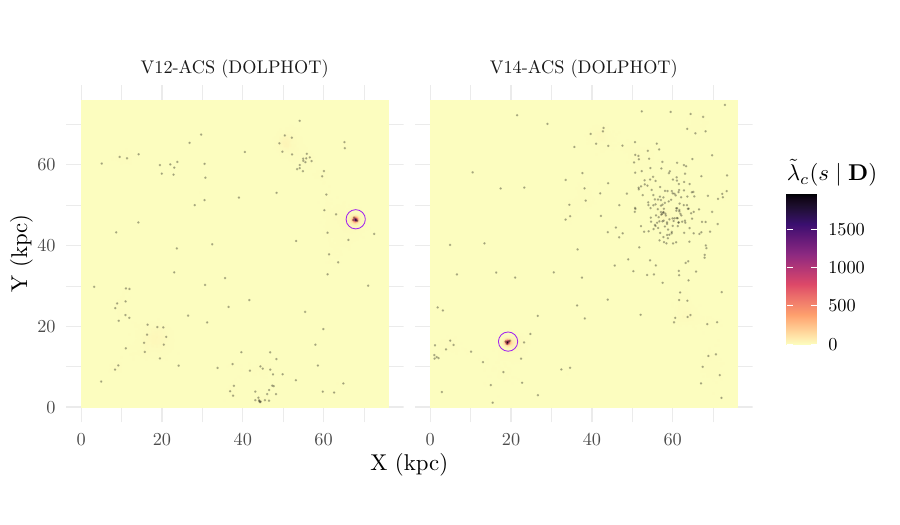}
    \caption{Scaled posterior probability (posterior/prior probability) of the potential locations of UDGs/dark galaxies in the images V12-ACS (left) and V14-ACS (right) obtained using the detection method in \cite{li2025poisson} based on the DOLPHOT GC data \citep{Harris2020}. Purple circles are the locations of CDG-2 in both images. Grey points are the locations of GC candidates.}
    \label{fig:CDG-2 detection DOL}
\end{figure*}

\section{Methods}\label{sec:methods}

We use the method of \cite{li2025poisson} for detecting UDGs/dark galaxies
based on the clustering of
GCs that do not belong to any apparent bright galaxies. 
The locations of GC candidates in an image are modeled as three overlapping point processes 
corresponding to GCs in the intergalactic medium (IGM), in luminous normal galaxies, and in UDGs/dark galaxies.  
The latent point pattern corresponding to the unknown UDG centers is the object of inferential interest.
Inference is conducted via a trans-dimensional Markov chain Monte Carlo (MCMC) algorithm where each MCMC sample is a point pattern of UDG centers. This allows
calculation of the posterior predictive probability that any region in an image contains a UDG.
New UDGs are detected by partitioning an image into a fine grid and plotting this probability on a color scale. 
A visual inspection is then performed to identify regions with relatively high probability of containing a
UDG/dark galaxy.

A significant advantage of the detection method by \cite{li2025poisson} is that the effects of background GCs are fully incorporated in the detection model. In fact, the method by \cite{li2025poisson} is able to produce posterior estimates on the probability that GC candidates in the image belong to a detected UDG.

\cite{li2025poisson} applied their method to GC data in 12 images from the DAOPHOT GC catalog \citep{Harris2020} to test their model performance. The model successfully detected all previously known UDGs with $N_{\mathrm{GC}} \geq 3$ in these 12 images as well as CDG-1. Additionally, a very strong clustering signal made up of three tightly clumped GCs was detected in both the images V12-ACS and V14-ACS, and this was the initial discovery of CDG-2. 

In the next section, we present the detection results and analysis of CDG-2 based on the new DOLPHOT GC catalog.

\section{CDG-2}\label{sec:res}

Figure \ref{fig:CDG-2 detection DOL} shows the detection results of CDG-2 using the DOLPHOT data. The purple circles in both panels of Figure \ref{fig:CDG-2 detection DOL} indicate the location of CDG-2 in V12-ACS and V14-ACS respectively. For better visualization and comparison, the color scale is the relative posterior probability of a UDG/dark galaxy present at a given location, i.e., it is the ratio of the posterior probability to the prior probability. For the prior probability, it is assumed that there are on average $2.5$ UDGs/dark galaxies in an image and that they are uniformly distributed across an image. From the posterior distribution, the probability that there is a UDG/dark galaxy at the location of CDG-2 is $\sim 2000$ times that of the prior probability.

On the other hand, based on results from \cite{li2025poisson} using the DAOPHOT data (see Figure 4 in \citealt{li2025poisson}), the scaled posterior intensity for CDG-2 is $\sim 200$ times that of the prior probability. The detection signal of CDG-2 thus increases by $10$-fold when using the new data from DOLPHOT.

The main reason that there is an increase in the detection signal of CDG-2 is the inclusion of the additional GC candidate CDG-2-GCC1. Adding one GC candidate in the close vicinity of the original three GC candidates in a region with relatively uncrowded background will drastically increase the intensity profile of CDG-2.
Another reason the detection signal is stronger under the new analysis is that we now consider and model the GC spatial distributions from all bright galaxies in both images, which reduces the noise affecting the detection of CDG-2. For simplicity, \cite{li2025poisson} did not include the modeling of GC distributions in these smaller galaxies, and their weaker GC clustering signals affected the signal strengths of CDG-2.

The four GC candidates in CDG-2 span a diameter of $3.2''$ which corresponds to $\sim 1.2$~kpc at $75$ Mpc. 
The probability that four GC candidates in CDG-2 occur due to random chance is $\sim 1.5\times10^{-5}$, or one in every $\sim67,000$ images with the same size and noise level as V12-ACS and V14-ACS, based on the posterior estimates of the background GC counts. 
In comparison, GC candidates in CDG-1 spans a diameter of $\sim 2$~kpc, while the same probability for CDG-1 under the DOLPHOT catalog is estimated to be $\sim 2\times 10^{-3}$. Thus, the GC candidates in CDG-2 are even more unlikely to arise randomly from the IGM than CDG-1 since the GCs in CDG-2 are much more compact.

In addition, for both V12-ACS and V14-ACS, the maximum a posteriori (MAP) estimate for the probability that all four GCs we analyze belong to a detected UDG (i.e., CDG-2) is $~94\pm{0.5}\%$. On the other hand, for all other GC candidates in both images, each of them has MAP estimate of $\lesssim 0.5\%$ chance that they belong to CDG-2. Therefore, we have strong statistical evidence (purely based on spatial distribution) that the four GCs indeed constitute the GC population of CDG-2 down to the detection limit.

\begin{figure*}[t]
    \centering
    \subfigure[]{\includegraphics[width = 0.8\textwidth]{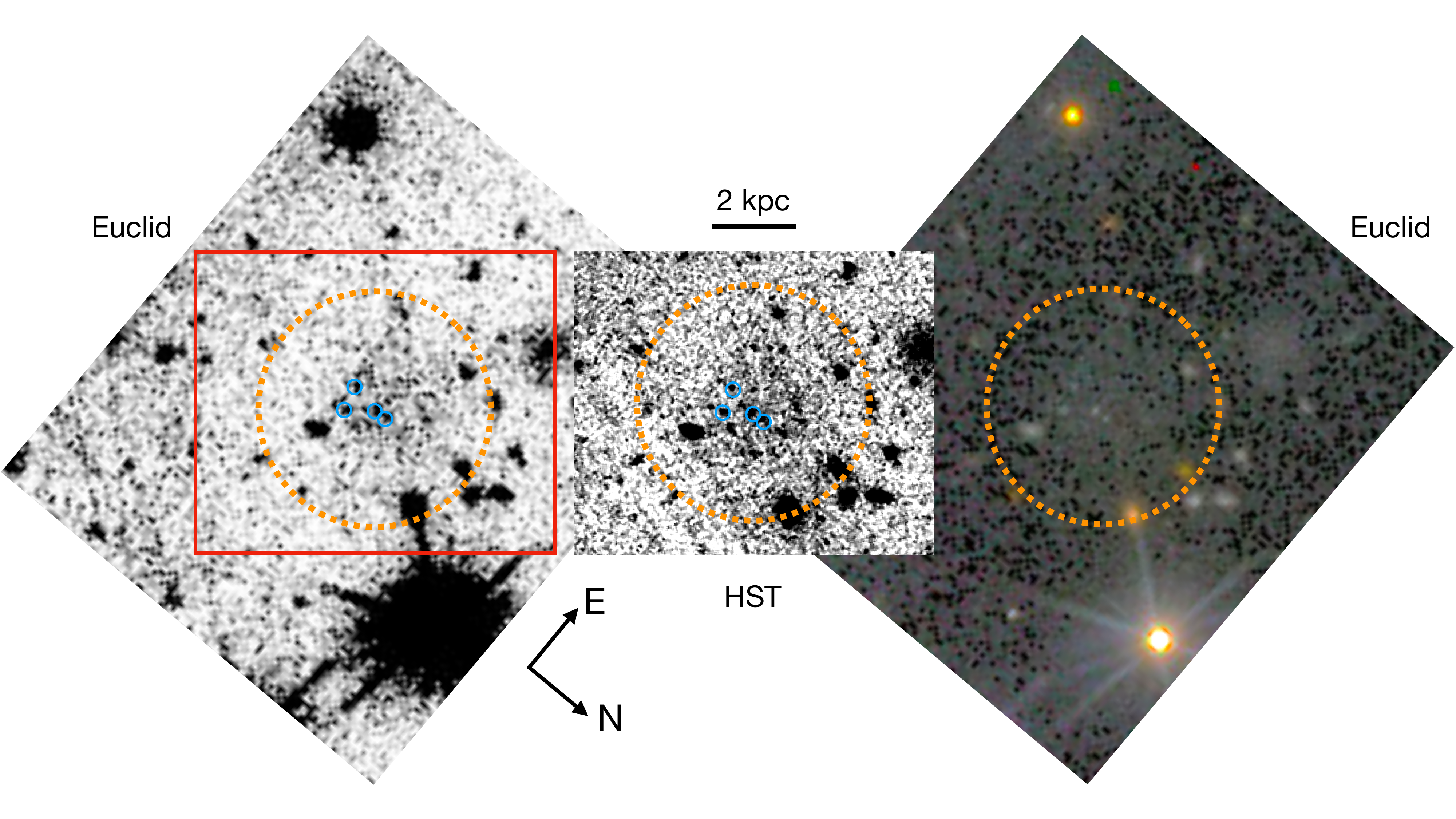}
    \label{fig:map}}
     \subfigure[]{\includegraphics[width = 0.98\textwidth]{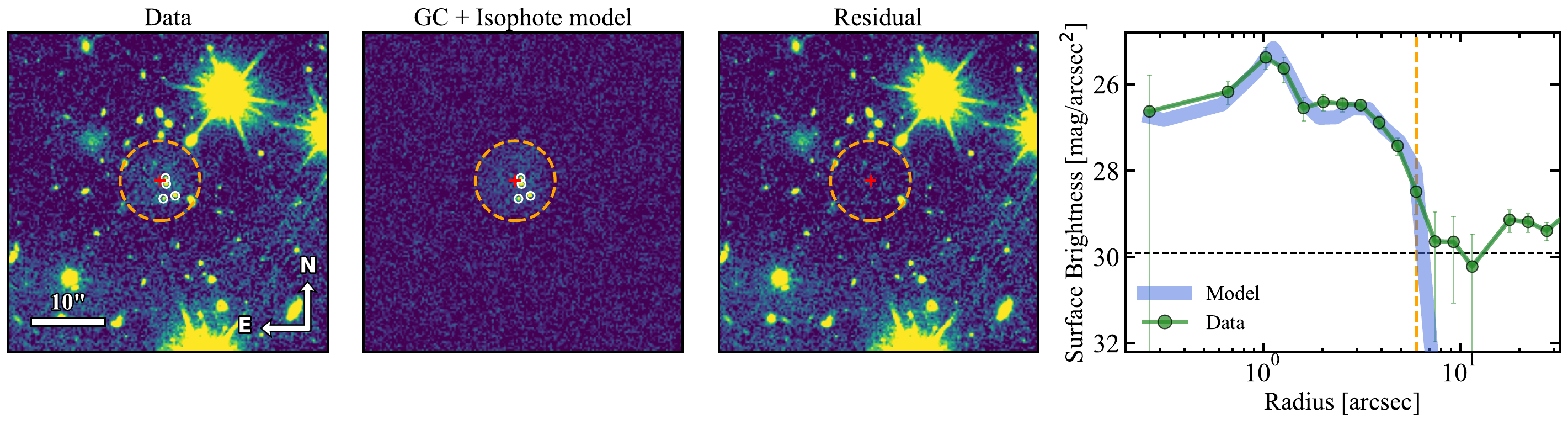}
    \label{fig:CDG-2 Euclid model}}
    \caption{(a) Cutout images for CDG-2 obtained from binning $I_E$-band (VIS) data from \textit{Euclid} (left); stacked and smoothed F814W images from V12-ACS and V14-ACS images from the PIPER \textit{HST} survey (middle); VIS, Y, and H band combined color image from \textit{Euclid} (right). Blue circles indicate the four GCs in CDG-2 while the orange dashed circle roughly outlines the region occupied by the diffuse emission in CDG-2. The diffuse emission in CDG-2 is clearly present in both data with similar morphology. (b) Left: {\it Euclid} $I_{E}$-band data used for modeling and analysis for CDG-2. The image scale is displayed in arcsinh stretch using [0.25, 0.95] quantiles of the image. Middle left: the combined model of GCs and diffuse emission. A simulated noise level matching the data is added. Middl right: residual of the image with the combined model subtracted. Right: radial profiles of the GC+isophote model (blue) and the data (green). The orange dashed line indicates the range of isophote fitting.
    In the left three panels, CDG-2 is marked with a orange circle and GCs are marked with white circles.}
\end{figure*}

\section{Follow-up Analysis}\label{sec:followup}

Given the high statistical significance that CDG-2 is not a random grouping of four GCs, we subsequently stacked the two images of V12-ACS and V14-ACS. The middle cutout image in Figure \ref{fig:map} shows the result: extremely diffuse emission surrounding the four GCs. 

Next, we checked for diffuse emission in CDG-2 using the newly released public data from the \textit{Euclid} ERO \citep{marleau2024}. The \textit{Euclid} data confirm that the diffuse emission is present, as shown in the left- and right-hand cutouts of Figure \ref{fig:map}. The morphology of the diffuse emission in both \textit{HST} and \textit{Euclid} data is almost identical. This means its presence is not due to imaging artifacts in either survey. The non-spurious diffuse emission coincident with extremely high GC clustering signal provides us with exceptionally strong evidence that CDG-2 is an almost dark galaxy. Certainly, the ultimate confirmation requires spectroscopy data obtained by powerful telescopes such as the \textit{James Webb Space Telescope}.

Another significant advantage of the \textit{Euclid} data over the \textit{HST} data from the PIPER survey is that \textit{Euclid} is optimized for detecting and analyzing diffuse structure in images \citep{cuillandre2024}. Thus, we used the single broad band $I_{E} \simeq r+i+z$ imaging data from the VIS imager onboard \textit{Euclid} to conduct a preliminary analysis of the diffuse emission and GC photometry. Note that the \textit{Euclid} data are in AB-mag instead of Vega-mag for the \textit{HST} data. Moreover, a $3\times3$-pixel binning was applied to the imaging data to improve diffuse emission analysis. We then obtained a rough estimate on the fraction of light contained in the GCs for CDG-2.

Specifically, we carried out combined modeling, using PSF photometry for the GCs and ellipse isophote fitting for the extended light, with the \texttt{photutils} software as follows: First, we removed the diffuse light in the background of CDG-2 using a \textsc{SExtractor}-like local background estimator and masked other nearby sources. CDG-2 was masked using an aperture with a radius of 6$\arcsec$. The background scale was chosen to be larger than CDG-2 to not remove the galaxy. Next, we ran PSF photometry on the background-subtracted image in which the target point sources (here the four GCs) were simultaneously fitted with the \textit{Euclid} PSF to obtain their integrated flux. The fitted GCs were subtracted from the image and the image was then smoothed by a Gaussian kernel with a standard deviation of 1 pixel to enhance the signal-to-noise ratio for diffuse light modeling. We then ran elliptical isophote fitting on the GC-subtracted, smoothed image. 

The above procedures were performed iteratively to mitigate the issue of point source modeling and diffuse light modeling affecting each other from one iteration to the other. For the isophote fitting, we set a maximum semi-major axis length of $6$~arcsec ($\sim 2.2$~kpc), beyond which the surface brightness of the isophote shows a sharp decrease. Since for low-surface brightness galaxies, the very faint outskirt regions can contain a significant fraction of their total flux \citep[e.g.,][]{Mihos_2015}, we also conducted a simple mock galaxy injection test to correct for the flux outside the range of the isophote model (see details in Appendix~\ref{appendix:mock_test}). We found that roughly $86\%$ of the diffuse light is contained within the isophote range. The total flux of the diffuse emission was first computed from the isophote model and then corrected based on the mock galaxy injection test. Due to the low-surface brightness of CDG-2, its perceived center is quite uncertain. Moreover, there is an offset between the center of the four GCs and the perceived center of the diffuse emission. Thus, we did not fix the centers of the isophotes but added constraints on their geometries to reject unrealistic results. Uncertainties in PSF photometry, background estimates, and isophote intensities were propagated to the calculation of the integrated flux of the GCs and the diffuse emission.

As mentioned in Section \ref{sec:methods} and \ref{sec:res}, the nature of our detection method and its results mean the effect of potential background GC contamination is of minimal concern. Therefore, it is safe to proceed our analysis by assuming the four GC candidates in CDG-2 are the contributors to its GC populations down to the detection limit of the images.

Figure~\ref{fig:CDG-2 Euclid model} shows the data and our analysis of isophote fitting for CDG-2. The combined GC and diffuse light model averaged over 100 iterations is shown in middle left panel of Figure~\ref{fig:CDG-2 Euclid model}, and the middle right panel shows the residual of the data and the model. The absolute AB magnitude of the four GCs is $M_{I_E, \mathrm{GC}}=-10.7\pm 0.21$~mag. After converting $I_E$-band magnitude to $V$-band magnitude\footnote{$M_{I_E} = M_V - 0.5$~mag.} \citep{marleau2024, saifollahi_2024}, the total GC luminosity is $L_{V, \mathrm{GC}} = 1.03\pm 0.2\times 10^6 L_{\odot}$ at a distance of 75 Mpc. Due to the extreme low surface brightness of CDG-2 ($\langle \mu\rangle_{I_{E}} \sim$ 27.0 mag/arcsec$^2$), the inferred absolute magnitude of the diffuse light within the isohpote fluctuates around $M_{I_E, \mathrm{diffuse}}^{\rm iso}=-12.3\pm 0.62$~mag, which leads to the diffuse light component of CDG-2 within the isophote range having a total luminosity of $L_{V, \mathrm{diffuse}}^{\rm iso} \approx 4.5\pm{2.5} \times 10^6 L_{\odot}$ at the same distance. 
The right panel of Figure~\ref{fig:CDG-2 Euclid model} shows the radial profiles of the combined GC and isophote model and the data using the light-weighted center of the isophote model as the center, indicating a good match between the data and the model.
The measured mean fraction of light in GCs within the isophote range stands at $18.2\pm2.0\%$.
The uncertainty combines the fluctuations over $100$ iterations and an average measurement uncertainty in each iteration. 
After correcting for the flux outside of the isophote range based on our injected mock galaxy test, the final estimate of the total luminosity for CDG-2 is $L_{V, \mathrm{gal}} \approx 6.2\pm{3.0} \times 10^6 L_{\odot}$. This translates to our final estimate for the mean GC light fraction of $f_{L, \rm GC} = 16.6 \pm {2.0}\%$ in CDG-2.

For robustness, we also considered the $20,000$~s Subaru Hyper Suprime-Cam $g$-band data \citep{Miyazaki_2018} and the archival CFHT/MegaCam $g$-band data to determine the GC light fractions in CDG-2. The modeling and data analysis procedures are similar to that for \textit{Euclid}. For simplicity, we do not illustrate the details here. We found that the resulting GC light fraction based on Subaru data is $\sim 18\%$, while the much noisier CFHT data yielded a GC light fraction of $\sim22\%$. The results here indicate that our estimates of the GC light fraction in CDG-2 are quite robust.

\section{Discussion}\label{sec:discussion}

\subsection{GC Stellar Mass Ratio}
As a first-order estimate, our results demonstrate that CDG-2 has some of the most extreme properties among all known galaxies. Based on the GC colors shown in Table \ref{tab:GC_candidates}, GCs in CDG-2 are rather blue and thus metal-poor, which is similar to GCs in NGC~5846-UDG1. If we make a similar assumption on the mass-to-light ratio as in \citet{Danieli2022} where $M/L_V \approx 2 M_\odot/L_\odot$ for field stars and $M/L_V
\approx 1.6M_\odot/L_\odot$ for GCs, we find an estimated GC stellar mass ratio of \textbf{$f_{M, \rm GC} \sim 13.7\%$}. However, the light and mass ratios are likely much higher since we did not correct for the GC luminosity function (GCLF). Indeed, based on the stacked \textit{HST} images in Figure~\ref{fig:map}, there does seem to be quite a few faint unresolved point-like sources around the four detected GC candidates that could potentially be undetected GCs.

Assuming CDG-2 has a GCLF similar to the canonical GCLF in $I$-band ($M_{\rm TO} = 26.3$~mag and $\sigma_M = 1.1$~mag), we here obtain a rough estimates on the GCLF-corrected GC-to-field star light and mass ratios. Based on the completeness fraction of DOLPHOT data \citep[see][for more details]{li2024mathpop}, the $50\%$ completeness limit is $25.75$~mag while the $90\%$ limit is $24.3$~mag. If we, for simplicity, assume the photometry is complete at the faintest magnitude of the four GCs ($25.2$~mag \footnote{The photometry is in fact $70\%$ complete at $25.2$~mag, so all the quoted ratio estimates for CDG-2 are underestimates.}) in Table \ref{tab:GC_candidates}, the cumulative GC luminosity at $25.2$~mag is $\sim 50\%$ of the total GC luminosity in CDG-2. This translates to a GC stellar light ratio of \textbf{$f_{L, \rm GC} \sim 33\%$}, while the GC stellar mass ratio is \textbf{$f_{M, \rm GC} \sim 28\%$}! Comparing this to known galaxies with extremely high $f_{M, \rm GC}$ estimates, such as NGC~5846-UDG1 at $\sim 9.8\%$ \citep{Danieli2022}, UGC 9050-Dw1 at $\sim 16\%$ \citep{fielder2023disturbed}, and VLSB-B at $\sim 23.7\%$ \citep{Gannon2024UDG}, CDG-2 may indeed be the galaxy with the highest GC-to-stellar mass ratio discovered to date.

Certainly, the above results heavily depend on the distance to CDG-2, $d_{\rm CDG-2}$, which causes changes in the GCLF. 
\begin{figure}[!t]
    \centering
    \includegraphics[width=0.5\textwidth]{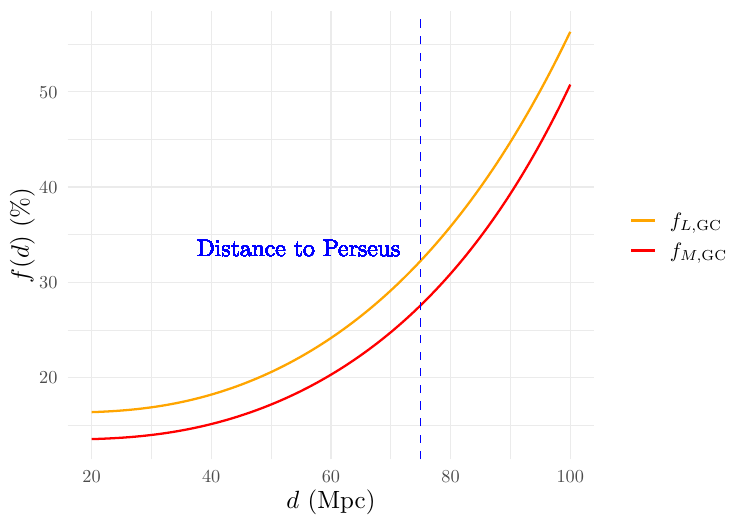}
    \caption{GCLF-corrected GC stellar light ($f_{L, \rm GC}$) and mass ($f_{M, \rm GC}$) ratios as a function of distance to CDG-2. The ratios here are corrected for potentially unobserved faint GCs by assuming a canonical GCLF adjusted for possible distance to CDG-2. For example, if the true distance to CDG-2 is $\sim20$~Mpc, the canonical GCLF would be similar to the magnitude distribution of the four observed GCs, and the GCLF-corrected ratios are the same as the calculated ratios based on the four GCs. Blue vertical dashed line indicates the distance of $75$~Mpc assumed for Perseus. }
    \label{fig:d-to-f}
\end{figure}
Figure \ref{fig:d-to-f} shows the GCLF-corrected $f_{L, \rm GC}$ and $f_{M, \rm GC}$ of CDG-2 as a function of $d_{\rm CDG-2}$. Under the canonical GCLF assumption, the result here shows that the GCLF-corrected light and mass ratios are respectively well above $23\%$ and $18\%$ even if CDG-2 is $20$~Mpc closer than previously assumed ($d_{\rm CDG-2} = 55$~Mpc).

\begin{figure*}
    \centering
    \includegraphics[width=0.9\linewidth]{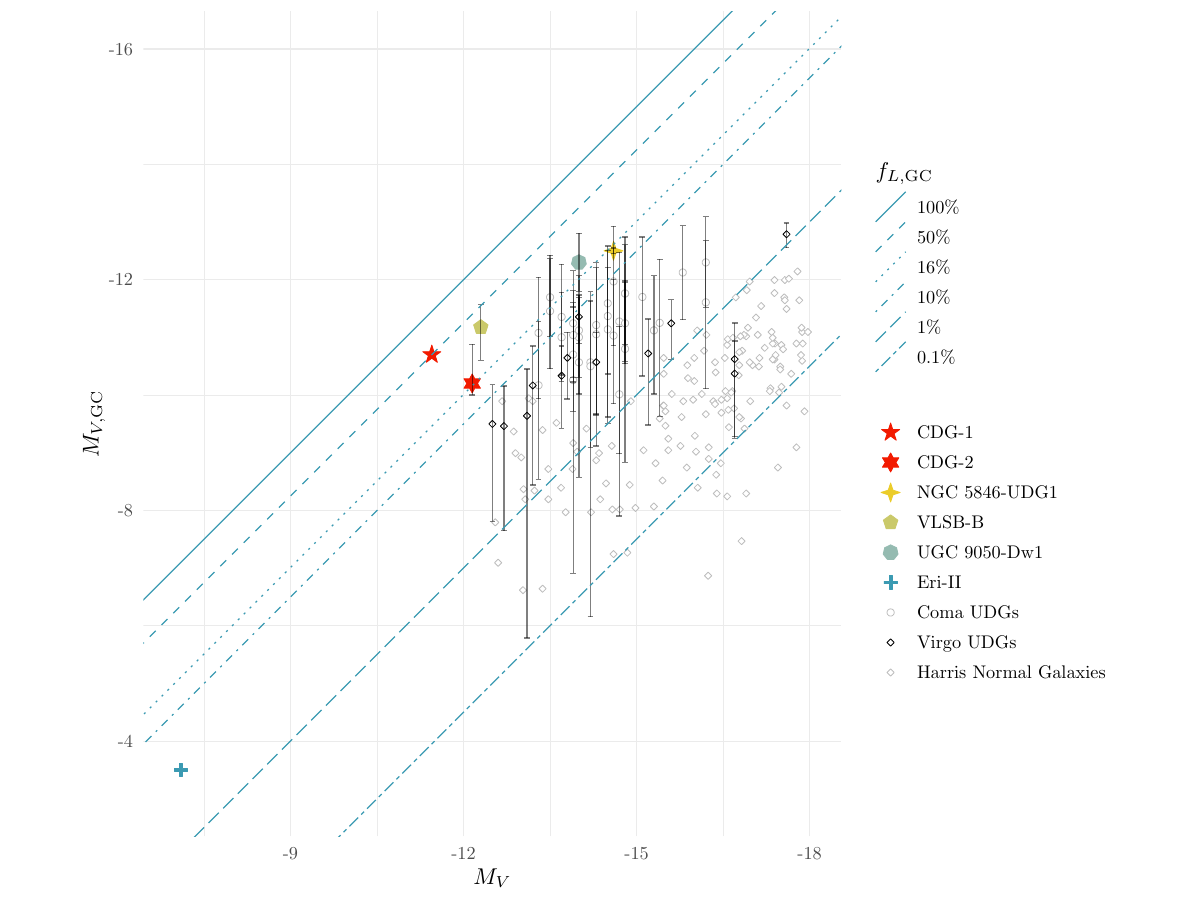}
    \caption{Galaxy total luminosity ($M_V$) as a function of GC system total luminosity ($M_{V, \rm GC}$) for various galaxies: Harris Normal Galaxies \citep{Harris+2013}; Coma UDGs \citep{Peng2016, van_Dokkum_2017, Lim2018}; Virgo UDGs \citep{Lim2020}; NGC~5846-UDG1 \citep{Danieli2022}; VLSB-B \citep{Lim2020, Toloba2023NGVS, Gannon2024UDG}; UGC~9050-Dw1 \citep{fielder2023disturbed}; Eri-II \citep{Crnojevic_2016}. Lines indicate various values of GC stellar light ratios. Note that the measurements for two dSph galaxies (KKs 55 and IKN) in the \cite{Harris+2013} catalog are updated based on \cite{Okamoto_2019, muller_2021}. Values for CDG-1 is taken from \cite{vanDokkum_2024} assuming its $f_{L, \rm GC} \geq 50\%$. Among all (confirmed) galaxies with GC populations, CDG-2 has one of the highest GC stellar light ratio while also being one of the faintest.}
    \label{fig:ratio_comp}
\end{figure*}

Moreover, the true value of $d_{\rm CDG-2}$ should be $\gtrsim 35$~Mpc. Otherwise, based on the measurements in Table \ref{tab:GC_candidates} and the completeness fraction for DOLPHOT, CDG-2 would become a system with a bottom-heavy GCLF. Such a system is rather unlikely since, to our best knowledge, it has never been found before. Given this constraint, the GCLF-corrected $f_{L, \rm GC}$ and $f_{M, \rm GC}$ should be at least $18\%$ and $15\%$ for CDG-2.

Despite all previous discussions, further observations are required to confirm the distance to CDG-2 as well as its GC population. For simplicity, we assume $d_{\rm CDG-2} = 75$~Mpc in all subsequent discussions unless specified otherwise. 

In addition, we only supply with strong confidence the estimates of $16.6\%$ and $13.7\%$ as lower bounds for  $f_{L, \rm GC}$ and $f_{M, \rm GC}$ of CDG-2. We intentionally restrain ourselves from concluding definitive estimates based on GCLF correction since it is well-known that there are significant variations in the GCLFs of dwarf-regime galaxies and UDGs \citep{Miller_2007, Jordan_2007, Villegas_2010, shen+2021, li2024mathpop}. Therefore, claims for the true ratios based on canonical GCLF and mere four GCs are highly uncertain.

Due to the distance to CDG-2 and the lower quality of data, we are not able to run a detailed GC evolutionary model as in \citet{Danieli2022} to infer the initial $f_{M, \rm GC}$ of CDG-2. However, given the current day lower-limit of $16.6\%$, the initial ratio should not be much different from unity as inferred for NGC~5846-UDG1. Therefore, CDG-2 can serve as a strong piece of evidence for the extreme star and galaxy formation scenario outlined in \cite{Trujillo_2021} and \cite{Danieli2022} where most of the star formation happened in extremely high gas surface density environment that produce massive GCs, while very little unbounded star formation occurred. 
In addition, the existence of CDG-2, along with other galaxies having high $f_{M, \rm GC}$, points to a potentially prevalent GC formation/destruction scenario of high GC formation efficiency and modest or very low rates of GC destruction \citep{Forbes_2025}.

The confirmation of CDG-2 also brings back CDG-1 into the conversation and may provide constraints on the nature of CDG-1. Even though previous observations \citep{Li2022, vanDokkum_2024, marleau2024} did not reveal detectable diffuse emission around CDG-1, the extremity of CDG-2 begs the question as to whether CDG-1 could be an even more extreme ``twin" of CDG-2 with hardly any stars formed outside of its GCs or that the GC populations were barely dissolved. 

We compare the properties of GC candidates in CDG-1 versus the ones in CDG-2 in Table \ref{tab:GC_candidates}. Interestingly, GCs in CDG-1 are on average brighter and bigger in size than those in CDG-2. The four GCs in CDG-1 are in total $\sim1.5\times$ brighter and twice as massive than those in CDG-2. This could hint at the possibility that CDG-1 is a copy of CDG-2 but at an earlier evolutionary stage where its GCs have not or barely dissolved, which would result in the appearance that CDG-1 does not have any detectable diffuse emission. 


Regardless of the possible formation mechanisms for CDG-1, if a $f_{L, \rm GC} \sim 33\%$ is possible for CDG-2, then it is not unreasonable to suspect that CDG-1 may be a galaxy with $f_{L, \rm GC} \gtrsim 50\%$ \citep{vanDokkum_2024}. In fact, based on the result in Figure \ref{fig:d-to-f}, it is entirely possible even for CDG-2 to achieve $f_{L, \rm GC} \gtrsim 50\%$ if it is at a slightly farther distance of $95$~Mpc. In addition, we plot in Figure \ref{fig:ratio_comp} the relationship between the total galaxy luminosity versus the GC system luminosity for various galaxies. CDG-1 and CDG-2 together with VLSB-B indeed seem to be in a league of their own given their high values of $f_{L, \rm GC}$ and extremely low luminosities. Further observations through radial velocity are certainly needed to provide more constraints on CDG-1.

\subsection{Overly Massive Dark Matter Halo}

In terms of the lowest surface brightness galaxies, the mean surface brightness for CDG-2 of $\langle\mu\rangle_V \sim 27.5$~mag arcsec$^{-2}$ ($\langle\mu\rangle_g \sim 27.8$~mag arcsec$^{-2}$), although a rough one, indicates that CDG-2 approaches the likes of RCP-32 \citep[$\langle\mu\rangle_g = 28.6$~mag arcsec$^{-2}$;][]{RCP_2021}. The difference is that the latter galaxy does not have a confirmed GC population as in CDG-2. In fact, CDG-2 is one of the lowest surface brightness galaxies with GC populations. As shown previously in Figure \ref{fig:ratio_comp}, CDG-2 clearly stands out in terms of having one of the highest value of $f_{L, \rm GC}$ while also being one of the faintest among the considered sample. The currently known faintest galaxy having a GC population is Eridanus II \citep{Crnojevic_2016}, with a central surface brightness of $\mu_{V,0} \sim 27.2$~mag arcsec$^{-2}$ and one single GC. However, the GC stellar light ratio as shown in Figure \ref{fig:ratio_comp} is at a much lower $4\%$ for Eridanus II. In contrast, even though many famous UDGs with significant GC populations, such as NGC~5846-UDG1, UGC~9050-Dw1, and some other UDGs in the Virgo and Coma cluster, may have a similar or potentially higher GC stellar light ratios, they are all much brighter than CDG-2 both in terms of the mean surface brightness and total luminosity. 

There has been increasing evidence that the number or the mass of GCs rather than the stellar mass are more indicative of the total galaxy halo mass \citep[e.g.,][]{Blakeslee1997, Peng_2008, Harris+2017, Burkert_Forbes2020, Lim2020, forbes2024ultra, Haacke_2025}. \cite{forbes2024ultra} found that UDGs with more GCs generally have higher halo masses and they do not follow the typical stellar-to-halo mass relations observed in normal galaxies. Therefore, the unique location CDG-2 occupies in Figure \ref{fig:ratio_comp} means that it could be an extreme case of galaxies with overly massive dark matter halos \citep{Beasley_2016, vanDokkum2016, forbes_2020, Lim2020, Toloba2023NGVS}, and it may well be the most extreme one known to date. 

Assuming CDG-2 has four GCs, its total stellar mass based on our previous estimations of luminosity and assumed mass-to-light ratio is $M_{*} \approx 1.2\times 10^7 M_{\odot}$, while the GC stellar mass is $M_{\rm GC} \approx 1.6 \times 10^6 M_\odot$. As a rough estimate, if the GC-to-halo mass relation of $M_{\rm GC}/M_h \approx 2.9\times 10^{-5}$ in \cite{Harris+2017} holds, then CDG-2 will have a halo mass of $M_h \approx  5.7 \times 10^{10} M_\odot$. Likewise, if the $M_h/M_\odot \approx 5\times 10^9 N_{\rm GC}$ scaling relation in \cite{Burkert_Forbes2020} holds for CDG-2, it would yield a similar total halo mass of $\sim 2\times 10^{10} M_{\odot}$. Both estimates put CDG-2 in the category of highly dark matter dominated galaxies \citep{VanDokkum2019, forbes_2020, Lim2020, Toloba2023NGVS, forbes2024ultra} with a halo mass fraction of $99.98\%$ and $99.94\%$ under the GC-to-halo mass relations of \cite{Harris+2017} and \cite{Burkert_Forbes2020}, respectively. However, this is once again not correcting for potentially unobserved GCs in CDG-2 for the GC-to-halo mass relation. If CDG-2 again has a canonical GCLF, its halo mass estimates are $\sim 1.2\times10^{11} M_\odot$ under both relations in \cite{Harris+2017} and \cite{Burkert_Forbes2020}, making its halo mass ratio $\gtrsim 99.99\%$. As a comparison, \cite{Toloba2023NGVS} used GC kinematics data and reported halo mass ratio estimates ranging from $99.5\%$ to $99.98\%$ within the half-number radii of GC systems for six Virgo UDGs \citep{Lim2020} with similar luminosity as CDG-2. In addition, \cite{Lim2020} considered photometric estimates for several Virgo UDGs with potentially extremely high dark matter fraction. Although \citet{Lim2020} did not provide a direct ratio estimates, the authors did report the GC specific frequency $S_N$ which can be illustrative of the extremity of CDG-2. Specifically, after correction by assuming the canonical GCLF, CDG-2 has a whopping specific frequency at $S_N = 345^{+323.1}_{-112.5}$, one of the highest ever recorded among all galaxies. Thus, the presence of unobserved GCs could imply that CDG-2 is even more extreme and may represent the example of a galaxy with the highest dark matter content. We do need to note that previously computed figures associated with CDG-2 rely heavily on assumptions and extrapolations of existing GC-to-halo mass relations.
It is thus essential to conduct further observations with high precision kinematic and spectroscopy data to constrain and confirm the dark matter content of CDG-2.

Along the same line of thought, further and higher-quality observations of CDG-1 are imperative since CDG-1 can turn out to be a galaxy that is even more extreme than CDG-2: it can be the first instance of a galaxy that is made up of pure dark matter halo without any field star population aside from a few GCs. The confirmed existence of two dark galaxies will provide an ideal testbed for models of fuzzy dark matter \citep{Hu2000, Walker2011, Hui2017}, which predict the existence of a highly compact soliton core in these dark matter halo galaxies \citep{Wasserman2019, Burkert2020}.

\subsection{Potentially Abnormal GC Populations}

On the flip side, if further observations reveal that CDG-2 does not contain other fainter GCs, then its constituent four GCs would make CDG-2 another galaxy with an abnormally bright GCLF similar to NGC~1052~DF2 and DF4 \citep{shen+2021}. In this case, CDG-2 would be another ideal candidate to study and test theories on clustered star formation and its potential implications. As proposed in \cite{vanDokkum2022}, the peculiar GC populations of NGC~1052~DF2 and DF4 may be due to a unique formation mechanism. However, more recent observations of galaxies with confirmed top-heavy bright GCLFs in various environments including DGSAT~I \citep{Janssens_2022}, FCC~224 \citep{Tang_2025}, R27, W88 \citep{li2024mathpop}, and possibly CDG-2 indicate that galaxies with top-heavy GCLFs may be more prevalent than once thought. 

\cite{shen+2021, li2024mathpop} hinted at the possibility that dwarf-regime galaxies and UDGs need not have a canonical GCLF and their GC populations could exhibit strong statistical variations caused by the highly varying star formation histories in their small halo. As more and more galaxies with top-heavy GCLFs and properties as extreme as CDG-2 are discovered, we have more opportunities and samples to constrain the theories of GCs and star formation.

%
\section*{Acknowledgments} D.L. would like to acknowledge funding support from Canadian Statistical Sciences Institute and Data Sciences Institute at the University of Toronto through grant number DSI-DSFY2R1P23. D.L. would also like to thank JiZhu Li for making sure that his eyes were not playing ticks on him while looking at CDG-2. W.E.H. would like to acknowledge funding support from NSERC. A.J.R. was supported by National Science Foundation grant AST-2308390. 

\section*{Data}

The HST data presented in this article were obtained from the Mikulski Archive for Space Telescopes (MAST) at the Space Telescope Science Institute. The specific observations analyzed can be accessed via \dataset[doi:10.17909/t87p-g529]{https://doi.org/10.17909/t87p-g529}.


\vspace{5mm}
\facilities{HST(ACS), HST(WFC3), Euclid}


\software{R, Python, photoutils}



\appendix
\section{Injection Test with a Mock Galaxy under \textit{Euclid}}
\label{appendix:mock_test}

\begin{figure}[!t]
    \centering
    \includegraphics[width=\textwidth]{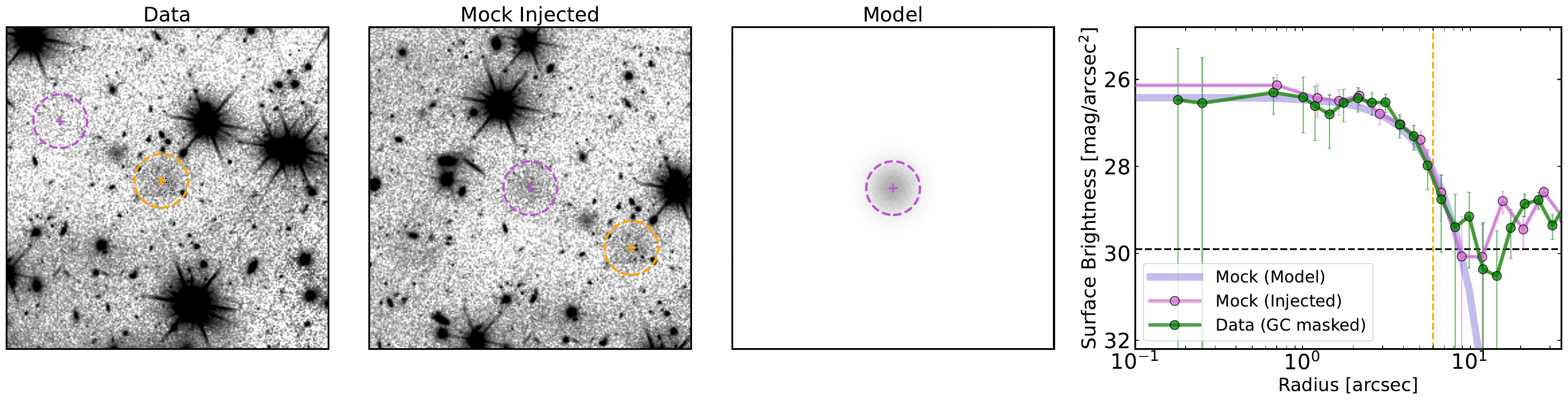}
    \caption{Injection test on the Euclid data with a mock UDG. Purple dashed circle indicates the isophote range of the injected mock UDG; Orange dashed circle indicates the isophote fitting range of CDG-2. Left: original Euclid data. Middle left: data with injected mock UDG. Middle right: mock UDG model. Right: radial profiles of CDG-2 with GC masked, the mock UDG model, and the injected UDG. The black dashed line indicates the 1-$\sigma$ depth of Euclid data on $10\arcsec \times10\arcsec $ scales. The orange dashed line indicates the range of isophote fitting.}
    \label{fig:CDG-2 mock}
\end{figure}

For the mock galaxy injection test, we considered injecting a mock UDG placed at the distance of the Perseus Cluster. Here we describe the injection procedure. The mock UDG is simulated using \texttt{ArtPop} \citep{greco2022artpop} by reproducing the surface brightness profile of CDG-2. \texttt{ArtPop} is a python software that generates artificial images of stellar systems with synthetic stellar populations. We created a mock UDG with a stellar mass of $M_* \sim 10^{7.4} M_{\odot}$ from the MIST isochrones \citep{Choi_2016, Dotter_2016} using a simple stellar population following a Kroupa initial mass function \citep{Kroupa_2001}. The mock UDG was set with a metallicity of $\mathrm{[Fe/H]}=-1$ and an age of $9$~Gyr. The galaxy was placed at $D=75$~Mpc and projected onto the image plane with the configuration of Euclid VIS, where stars were sampled following the distribution of a 2D S{\'e}rsic profile with a S{\'e}rsic index $n_{\rm s\acute{e}rsic}=0.5$, an ellipticity $\epsilon=0.05$, and an apparent effective radius $R_{\rm eff}$ = 1.5 kpc. The injection process and the radial profiles of data (with GC masked), mock model, and data with injection are illustrated in Figure \ref{fig:CDG-2 mock}, indicating a reasonable match between the mock galaxy to the CDG-2. However, it should be noted that the parameters chosen is only for reproducing the light distribution and thus do not represent real properties of CDG-2.

We then performed isophote fitting on the injected mock UDG with the same procedures and setups as CDG-2. 
The total flux of the isophote model includes $86\%$ of the total flux of the injected mock UDG, indicating a low-level light loss in the very faint outskirts thanks to the sharp extended PSF wing of Euclid VIS \citep{cuillandre2024}. We corrected the diffuse light flux and accordingly, the flux ratio, using this factor. We noticed that the fraction of light in the outskirts of the galaxy that needs to be corrected is dependent on the intrinsic physical parameters ($M_*$, $R_{\rm eff}$, $n_{\rm sersic}$) of the UDG. However, the integrated flux after correction is similar, i.e., a UDG with a shallower, extended profile yields a larger correction, while a larger UDG with a steeper outskirts has a smaller correction. Therefore, such degeneracy in the physical parameters of the UDG would not dramatically affect the total flux and the GC-to-total flux ratio after correction.

\bibliography{bibliography}{}
\bibliographystyle{aasjournal}


\end{CJK}
\end{document}